\documentclass{elsart}

\usepackage{graphicx,natbib,amssymb,lineno,hyperref}
\usepackage{lscape}

\def\astrobj#1{#1}

\newcommand\apj{{ApJ}}%
\newcommand\aj{{AJ}}%

\begin{document}

\begin{frontmatter}
\title{First analysis of eight Algol-type binaries: EI~Aur, XY~Dra, BP~Dra, DD~Her, VX~Lac, WX~Lib, RZ~Lyn, and TY~Tri}

\author{P. Zasche}
\ead{zasche@sirrah.troja.mff.cuni.cz}

\address{Astronomical Institute, Faculty of Mathematics and Physics,
 Charles University in Prague, CZ-180 00 Praha 8, V Hole\v{s}ovi\v{c}k\'ach 2, Czech Republic}

\begin{abstract}
The available photometry from the online databases were used for the first light curve analysis of
eight eclipsing binary systems \astrobj{EI Aur}, \astrobj{XY Dra}, \astrobj{BP Dra}, \astrobj{DD
Her}, \astrobj{VX Lac}, \astrobj{WX Lib}, \astrobj{RZ Lyn}, and \astrobj{TY Tri}. All these stars
are of Algol-type, having the detached components and the orbital periods from 0.92 to 6.8~days.
For the systems EI~Aur and BP~Dra the large amount of the third light was detected during the light
curve solution. Moreover, 468 new times of minima for these binaries were derived, trying to
identify the period variations. For the systems XY~Dra and VX~Lac the third bodies were detected
with the periods 17.7, and 49.3~yr, respectively.
\end{abstract}


\end{frontmatter}

\section{Introduction}

The role of eclipsing binaries in nowadays astrophysics is undisputable. We use the eclipsing
binary systems (hereafter EB)  for the most accurate determination of the stellar masses, radii, as
distance indicators, or as classical celestial mechanics laboratories. We can test the stellar
structure models even outside of our Galaxy, see e.g. \cite{2004NewAR..48..731R}. Additionally,
also the hidden components can be studied via the dedicated observations of particular binaries as
well as the dynamical effects in such multiple systems \citep{2013ApJ...768...33R}. Due to all of
these reasons the photometric monitoring and analysis of the light curves of selected eclipsing
binaries still presents a fruitful contribution to the stellar astrophysics.

On the other hand, the available photometry for many interesting eclipsing binaries exists, but
some of these EBs were still not analysed yet. Hence, we decided to use mainly the Super WASP
photometry \citep{2006PASP..118.1407P} for a light curve analysis and derivation of new minima
times for such systems, which were not studied before and their light curve solution is missing.

\section{Analysis}

The selection criteria for the binaries included in our study were the following. Only such
binaries with known orbital periods were chosen, having no light curve solution published up to
date, have enough data points for the analysis and also have several published times of minima. The
last point was checked via an online archive of minima times observations, a so-called $O-C$
gateway\footnote{http://var.astro.cz/ocgate/} \citep{2006OEJV...23...13P}. Due to the very good
time coverage provided by the Super WASP survey we used this database for the whole analysis of the
light curve. The other databases such as NSVS \citep{2004AJ....127.2436W}, ASAS
\citep{2002AcA....52..397P}, CRTS \citep{2009ApJ...696..870D}, or OMC \citep{OMC} were used only
for deriving the times of minima for a subsequent period analysis. All of the studied systems are
the northern-hemisphere stars of moderate brightness (10~mag $<$ V $<$ 15~mag) and with the orbital
periods ranging from 0.9 to 6.8 days.

For analysing the light curves we used the {\sc PHOEBE} program \citep{Prsa2005}, which is based on
the algorithm by \cite{Wilson1971}. Having sometimes rather limited information about the stars,
some of the parameters have to be fixed for the light curve (hereafter LC) solution. At first, the
"Detached binary" mode (in Wilson \& Devinney mode 2) was assumed for computing. The value of the
mass ratio $q$ was set to 1. The limb-darkening coefficients were interpolated from van~Hamme's
tables (see \citealt{vanHamme1993}), and the linear cosine law was used. The values of the gravity
brightening and bolometric albedo coefficients were set at their suggested values for
convective or radiative atmospheres (see \citealt{Lucy1968}). 
Therefore, the quantities which could be directly calculated from the LC are the following: the
relative luminosities $L_i$, the temperature of the secondary $T_2$, the inclination $i$, and the
Kopal's modified potentials $\Omega_1$ and $\Omega_2$. The synchronicity parameters $F_1$ and $F_2$
were also fixed at values of 1. The value of the third light $L_3$ was also computed if a
non-negligible value resulted from the fitting process. And finally, the linear ephemerides were
calculated using the available minima times for a particular system.

With the final LC analysis, we also derived many times of minima for a particular system, using a
method as presented in \cite{2014A&A...572A..71Z}. The template of the LC was used to fit the
photometric data from the Super WASP as well as from other surveys. This set of minima times was
then combined with the already published minima mostly taken from the $O-C$ gateway
\citep{2006OEJV...23...13P}.

\section{The individual systems}

\subsection{EI Aur}

The system EI Aur (also GSC 02392-00102 ) was discovered by \cite{1936AN....259...37H}, who also
classified the star as an Algol-type. Its orbital period is of about 1.2~days, but there were no
light curve or any spectroscopic analysis performed. We can only roughly estimate its type from the
color indices, hence we fixed the primary temperature at a value of 6000~K for the whole fitting
process.

\begin{figure}
 \includegraphics[width=12cm]{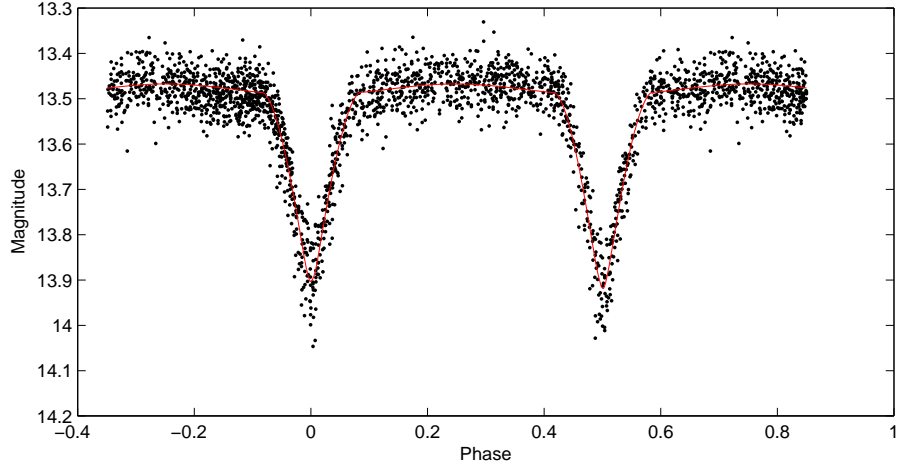}
 \caption{Light curve analysis of EI~Aur, based on the Super WASP photometry.}
 \label{FigEIAurLC}
\end{figure}

The Super WASP photometry revealed that it is a detached system, having both minima of roughly
equal depths. Therefore, the {\sc PHOEBE} code was used to these data and the LC fit is presented
in Fig. \ref{FigEIAurLC}, while the LC parameters are given in Table \ref{TableLC}. As one can see,
the secondary component has almost the same temperature as the primary, hence there is still a
doubt which of the minima is the primary one. Another interesting finding is the fact that
relatively large contribution of the third light was detected in the LC solution. This would
naturally explain why both the minima have only so shallow depths.

\begin{figure}
 \includegraphics[width=12cm]{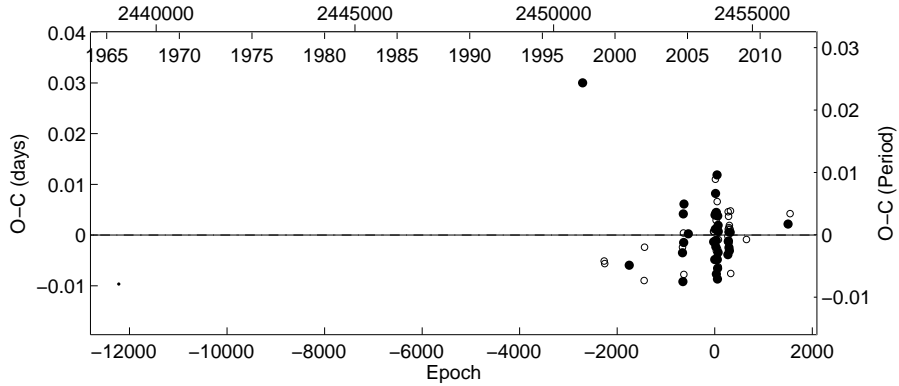}
 \caption{O-C diagram of times of minima derived from available photometry for EI~Aur. The black points stand for the
 primary minima, while the open circles stand for the secondary ones. The larger the symbol, the higher the weight.}
 \label{FigEIAurOC}
\end{figure}

\begin{table*}[t]
 \tiny
 \caption{The light-curve parameters as derived from our analysis.}
 \label{TableLC} \centering
\begin{tabular}{ c c c c c c c}
\hline \hline
 Parameter     &    EI Aur                &    XY Dra               &      BP~Dra               &    DD~Her                 \\ \hline
 $JD_0-2400000$& 54050.6460 $\pm$ 0.0019  & 54597.5551 $\pm$ 0.0160 & 54659.5430 $\pm$ 0.0013   & 53165.3571 $\pm$ 0.0152   \\
 $P$ [d]       &1.2266930 $\pm$ 0.0000013&2.3152311 $\pm$ 0.0000342& 0.9868093 $\pm$ 0.0000003 & 5.6433970 $\pm$ 0.0000061 \\
 $i$ [deg]     & 87.67 $\pm$ 0.59         & 89.76 $\pm$ 0.40        &  86.97 $\pm$ 0.52         &  79.61 $\pm$ 0.35         \\
 $T_1$ [K]     & 6000 (fixed)             & 6500 (fixed)            &  6000 (fixed)             &  8800 (fixed)             \\
 $T_2$ [K]     & 6044 $\pm$ 38            & 4343 $\pm$  34          &  5429 $\pm$  41           &  4985 $\pm$ 50            \\
 $\Omega_1$    & 4.847 $\pm$ 0.059        & 5.942 $\pm$ 0.039       & 5.312 $\pm$ 0.052         &  5.354 $\pm$ 0.022        \\
 $\Omega_2$    & 5.181 $\pm$ 0.072        & 4.619 $\pm$ 0.020       & 6.216 $\pm$ 0.063         &  10.671 $\pm$ 0.021       \\
 $L_1$ [\%]    & 38.3 $\pm$ 0.5           & 83.7 $\pm$ 0.8          & 45.3 $\pm$ 0.9            &  93.7 $\pm$ 0.8           \\
 $L_2$ [\%]    & 32.9 $\pm$ 0.7           & 16.3 $\pm$ 0.5          & 18.5 $\pm$ 0.8            &   6.3 $\pm$ 0.6           \\
 $L_3$ [\%]    & 28.8 $\pm$ 0.7           &  0.0 $\pm$ 0.0          & 36.2 $\pm$ 0.4            &   0.0 $\pm$ 0.0           \\ \hline
\end{tabular}
\end{table*}

One can ask whether the third body detected in the LC solution is somehow gravitationally bounded
with the eclipsing pair or is it just a coincidence as a so-called optical binary. We also derived
the times of minima from the Super WASP photometry and plotted them together with the already
published ones in Fig. \ref{FigEIAurOC}. As one can see, there is no obvious variation in the times
of minima. The data coverage is still rather poor, but even this diagram can be used to set some
tighter limits for the parameters of the proposed additional body in the system.

\subsection{XY Dra}

Eclipsing binary called XY Dra (also AN 391.1929) was discovered as a variable by
\cite{1929AN....236..245W}. Since then no detailed analysis of this star was carried out, only a
few minima times were published, mostly in the last 20 years. Its spectral type is not known, hence
we fixed the primary temperature at a value of 6500~K in agreement with the photometric indices as
published in different databases.

\begin{figure}
 \includegraphics[width=12cm]{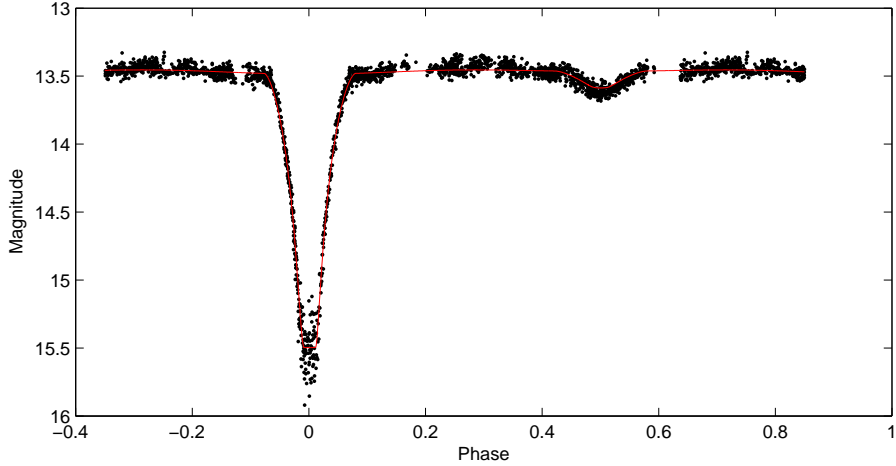}
 \caption{Light curve analysis of XY~Dra, based on the Super WASP photometry.}
 \label{FigXYDraLC}
\end{figure}

The light curve solution was carried out in the same way as for the previous system, resulting in
light curve parameters given in Table \ref{TableLC}, while the fit is presented in Fig.
\ref{FigXYDraLC}. As one can see, the primary component is very dominant star in the system and no
third light was detected. On the other hand, there still remains an open question whether the data
points located in the center of the primary eclipse are real or not. One possible explanation is
that there is a total eclipse (as assumed in our fit), or the second explanation could be that we
just reached a limit for the Super WASP photometry of about 15.5~mag and the real eclipse is much
deeper. Only further dedicated observations of the primary eclipse would reveal its true nature.

\begin{figure}
 \includegraphics[width=12cm]{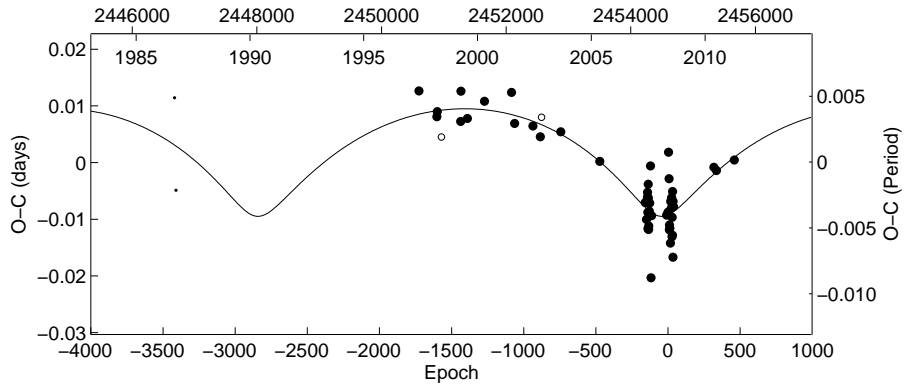}
 \caption{O-C diagram of times of minima derived from available photometry for XY~Dra. The solid curve represents
 the final fit (see the text for details).}
 \label{FigXYDraOC}
\end{figure}

Moreover, we also derived the times of minima from the Super WASP photometry and combined them with
the already published more precise data from the "O-C gateway". The result of our fitting is
presented in Fig. \ref{FigXYDraOC}. We used a so-called light-travel-time effect for describing the
variation in the $O-C$ diagram. This method was described elsewhere, e.g. \cite{Irwin1959} or
\cite{Mayer1990}, using a set of five orbital parameters for description of the third body orbit
around the eclipsing binary itself. With this assumption we found that the variation with period of
about 17.7~yr, eccentricity 0.59 and a semiamplitude of 0.0095~days is probably caused by a third
body of such a small mass that its light contribution to the total light of the system would be
negligible. This would be a reason why no third light was detected in the LC solution as presented
in Table \ref{TableLC}.

\subsection{BP~Dra}

The star BP~Dra (also GSC 04231-01877) was discovered by \cite{1966VeSon...7...61G}. It has the
orbital period shorter than one day, of about 0.99~days, which makes it rather hard to observe for
the minima timings. This is maybe a reason why only so little information is known about this star.
Lacking enough data, we have to fix the primary temperature at a value of 6000~K in agreement with
the color indices (\citealt{2006AJ....131.1163S} and \citealt{2010PASP..122.1437P}).

\begin{figure}
 \includegraphics[width=12cm]{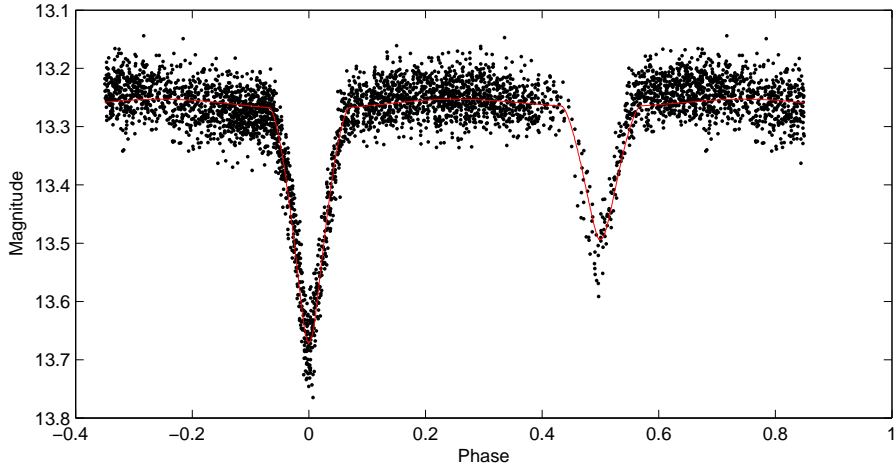}
 \caption{Light curve analysis of BP~Dra, based on the Super WASP photometry.}
 \label{FigBPDraLC}
\end{figure}

The LC solution was found and the parameters are given in Table \ref{TableLC}, while the plot with
the final fit is presented in Fig. \ref{FigBPDraLC}. As one can see, both eclipses are rather
shallow and the scatter of the observations is quite large. However, this is due to the fact that
there was detected also a non-negligible amount of the third light in the LC solution. We can only
speculate about its origin and classify the system as a potential triple.

\begin{figure}
 \includegraphics[width=12cm]{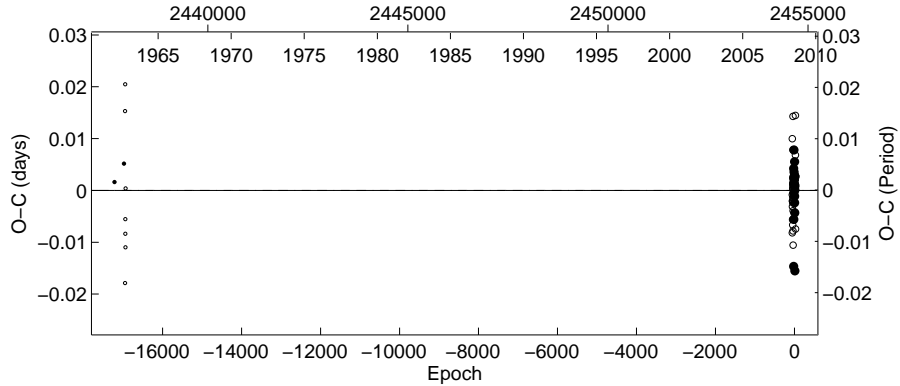}
 \caption{O-C diagram of times of minima for BP~Dra.}
 \label{FigBPDraOC}
\end{figure}

Collecting all available Super WASP photometry we also derived about 50 new minima timings. These
data are plotted together with the ones from 1960's and indicate that there is obviously no period
change during these fifty years. Only a more detailed analysis would reveal the third-body
hypothesis as a plausible or implausible.

\subsection{DD~Her}

Another system in our sample of stars is DD~Her (also TYC 2103-352-1), discovered as a variable by
\cite{1929AN....236..233H}. It has relatively longer period of about 5.6~days, however it is also a
system lacking of any detailed analysis. Due to its spectral classification as A2, we used the
primary temperature of 8800~K for the whole LC fitting process.

\begin{figure}
 \includegraphics[width=12cm]{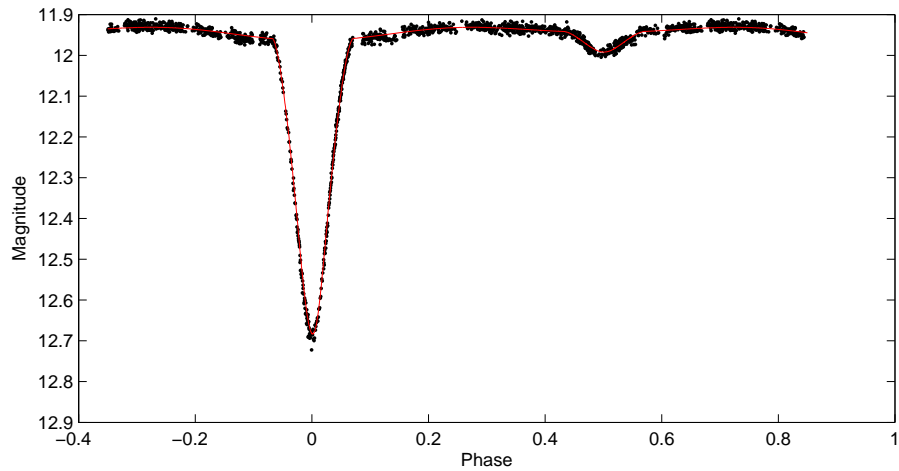}
 \caption{Light curve analysis of DD~Her, based on the Super WASP photometry.}
 \label{FigDDHerLC}
\end{figure}

\begin{figure}
 \includegraphics[width=12cm]{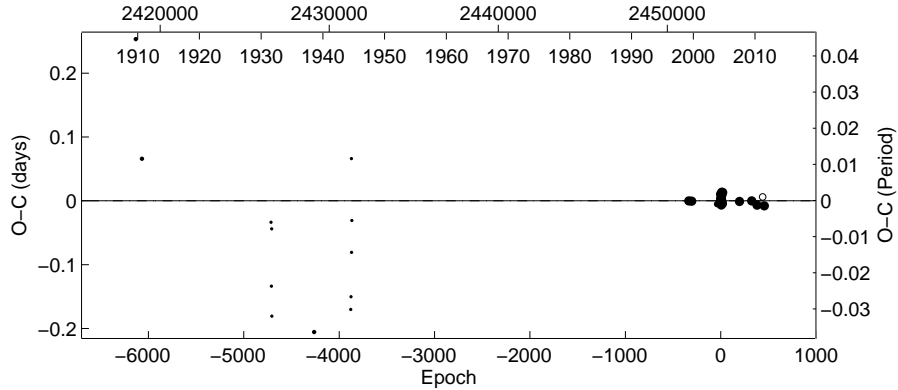}
 \caption{O-C diagram of times of minima for DD~Her.}
 \label{FigDDHerOC}
\end{figure}

The Super WASP photometry used for the LC analysis yielded a LC solution given in Table
\ref{TableLC}, while the LC fit is plotted in Fig. \ref{FigDDHerLC}. As one can see, the primary
star is absolutely dominant in the system and no third light was detected. Due to its deep and
well-covered eclipses we derived only the primary minima for a period analysis and the $O-C$
diagram (plotted in Fig. \ref{FigDDHerOC}). Also the ASAS and NSVS photometry was used for deriving
two minima times. The older observations suffer from large scatter and are almost useless for any
analysis. More recent data points show no variation.

\subsection{VX~Lac}

The system VX~Lac (also TYC 3214-1295-1) is probably the most studied star in our sample of
binaries. \cite{Cannon1934} derived its spectral type as F0, and since then no other more recent
classification was carried out. The system was also included into the study of eclipsing binaries
with period changes and third bodies by \cite{2008NewA...13..405Z}, who discovered a variation of
about 68~yr and 0.02~days semiamplitude.

\begin{figure}
 \includegraphics[width=12cm]{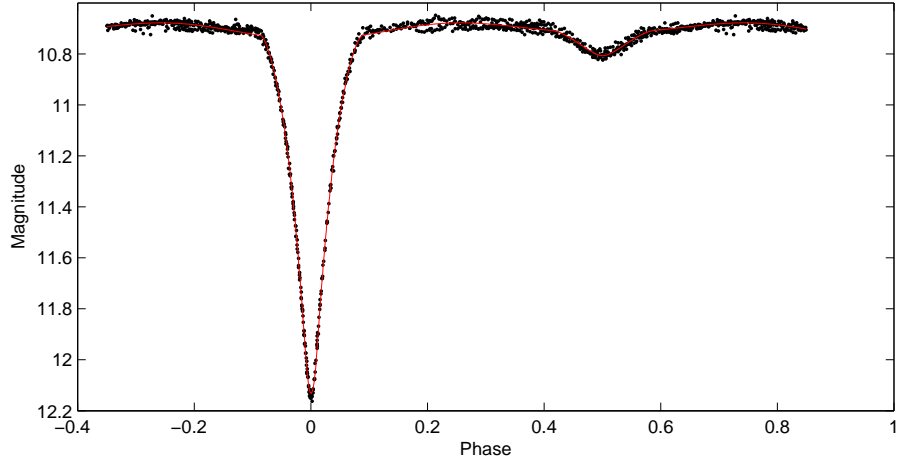}
 \caption{Light curve analysis of VX~Lac, based on the Super WASP photometry.}
 \label{FigVXLacLC}
\end{figure}

We collected the Super WASP photometry for a LC solution and analysed the system using the {\sc
PHOEBE} program. The primary temperature was set to 7228~K using the F0 spectral classification and
also the Tycho-2 data from \cite{2006ApJ...638.1004A}.  The result is plotted in Fig.
\ref{FigVXLacLC} and the LC parameters are given in Table \ref{TableLC2}. As one can see, the
primary is the dominant component in the system. No third light was detected, which sets some
constraints on the third-body hypothesis as resulted from the period analysis. There was also found
that either of the stars is probably slightly physically variable, because in the outside-eclipse
region there were seen some period-to-period deviations, which cause a slightly larger scatter of
the data near quadratures.

\begin{figure}
 \includegraphics[width=12cm]{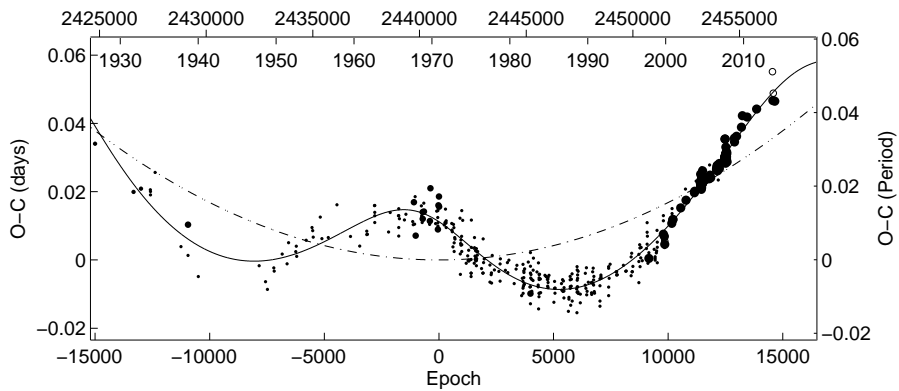}
 \caption{O-C diagram of times of minima for VX~Lac. The solid line represents the final fit,
 while the dash-dotted line stands for the quadratic ephemerides term.}
 \label{FigVXLacOC}
\end{figure}

The analysis of period was performed on the already published data as well as our new derived times
of minima from the Super WASP and NSVS photometry. Our updated solution as presented in Fig.
\ref{FigVXLacOC} represents only a slight correction of the already published one by
\cite{2008NewA...13..405Z}. New values of period, eccentricity and semiamplitude are: 49.3~yr,
0.239, and 0.0144~days, respectively. However, even such a result is able to explain a
non-detection of the third light in the LC solution. The potential third body is probably so small
(hence has so low luminosity) that it cannot be detected in the LC solution.

\begin{table*}[t]
 \tiny
 \caption{The light-curve parameters as derived from our analysis.}
 \label{TableLC2} \centering
\begin{tabular}{ c c c c c c c}
\hline \hline
 Parameter     &    VX~Lac                &    WX~Lib               &      RZ~Lyn               &    TY~Tri                 \\ \hline
 $JD_0-2400000$& 40908.9074 $\pm$ 0.0012  & 54564.5560 $\pm$ 0.0007 & 45347.3639 $\pm$ 0.0056   & 35778.5110 $\pm$ 0.0658   \\
 $P$ [d]       &1.07449709 $\pm$ 0.0000014&0.9200078 $\pm$ 0.0000003& 1.1469092 $\pm$ 0.0000007 & 6.7620349 $\pm$ 0.0000381 \\
 $i$ [deg]     & 86.84 $\pm$ 0.25         & 85.05 $\pm$ 0.17        &  76.03 $\pm$ 0.36         &  89.94 $\pm$ 0.27         \\
 $T_1$ [K]     & 7228 (fixed)             & 5200 (fixed)            &  8800 (fixed)             &  5720 (fixed)             \\
 $T_2$ [K]     & 4486 $\pm$ 36            & 5035 $\pm$ 27           &  6119 $\pm$ 62            &  5487 $\pm$ 40            \\
 $\Omega_1$    & 4.589 $\pm$ 0.019        & 4.808 $\pm$ 0.023       & 3.961 $\pm$ 0.025         &  8.871 $\pm$ 0.032        \\
 $\Omega_2$    & 5.021 $\pm$ 0.018        & 5.728 $\pm$ 0.029       & 4.295 $\pm$ 0.046         &  12.906 $\pm$ 0.041       \\
 $L_1$ [\%]    & 94.1 $\pm$ 0.9           & 64.3 $\pm$ 1.0          & 83.8 $\pm$ 0.8            &  74.0 $\pm$ 0.7           \\
 $L_2$ [\%]    &  5.9 $\pm$ 0.8           & 35.5 $\pm$ 0.9          & 13.4 $\pm$ 0.7            &  26.0 $\pm$ 0.3           \\
 $L_3$ [\%]    &  0.0 $\pm$ 0.0           &  0.2 $\pm$ 0.4          &  2.8 $\pm$ 0.4            &   0.0 $\pm$ 0.0           \\ \hline
\end{tabular}
\end{table*}

\subsection{WX~Lib}

The star WX~Lib is relatively seldom studied system. It is the southernmost star in our sample,
hence it was also observed by the ASAS survey. It is also the star with the shortest orbital period
in our sample, of about 0.92~days only (while the GCVS and VSX still presents the period
0.46~days).

\begin{figure}
 \includegraphics[width=12cm]{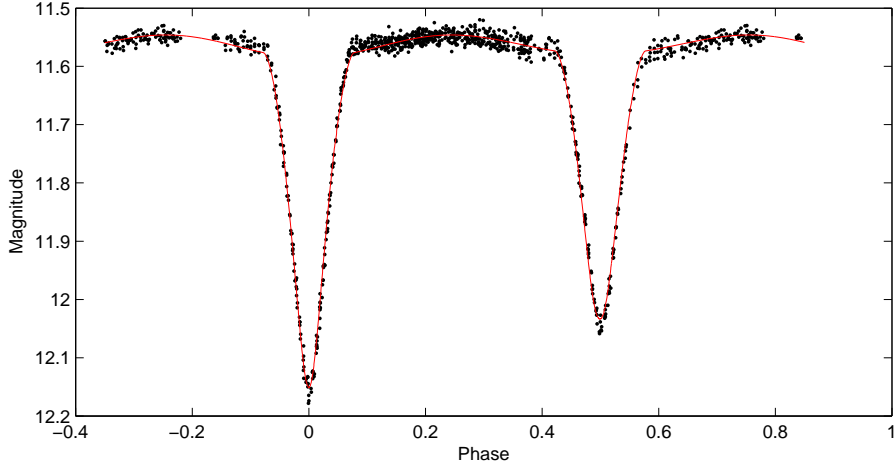}
 \caption{Light curve analysis of WX~Lib, based on the Super WASP photometry.}
 \label{FigWXLibLC}
\end{figure}

Due to lacking relevant information about the system, we fixed the primary temperature to the value
of 5200~K in agreement with the photometric indices found in various databases. The LC shows two
rather similar and relatively deep eclipses. The LC solution (see Fig. \ref{FigWXLibLC} and Table
\ref{TableLC2}) reveals that both stars are somehow similar, with the primary slightly dominating.
The value of the third light remains negligible.

\begin{figure}
 \includegraphics[width=12cm]{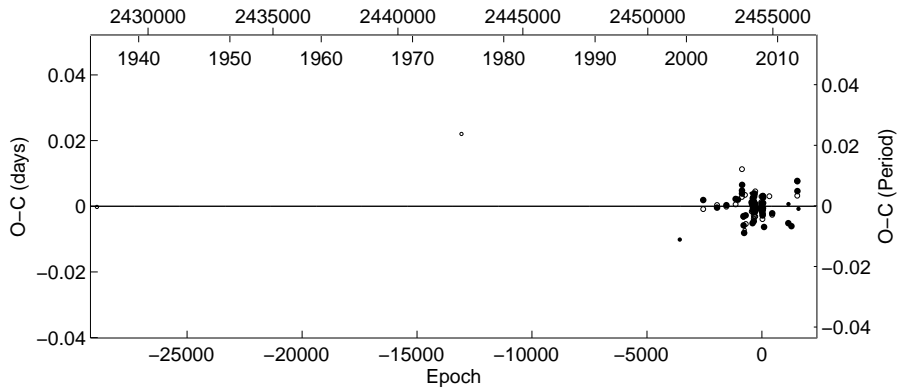}
 \caption{O-C diagram of times of minima for WX~Lib.}
 \label{FigWXLibOC}
\end{figure}

The period variation was studied using the Super WASP as well as the ASAS, CRTS, and OMC
photometric data. Because of only very poor coverage of the time interval since its discovery to
the nowadays time we are not able to identify any period variation in this time interval.

\subsection{RZ~Lyn}

RZ Lyn (also GSC 02995-00972) is another rather neglected eclipsing binary, briefly studied by
\cite{1953AN....281..183H}, who classified it as a $\beta$~Lyrae type star and giving its correct
orbital period of about 1.147~days. Its spectral of A2 type was given by
\cite{1961MitVS.569....1G}. Since then only several publications with the times of minima were
published.

\begin{figure}
 \includegraphics[width=12cm]{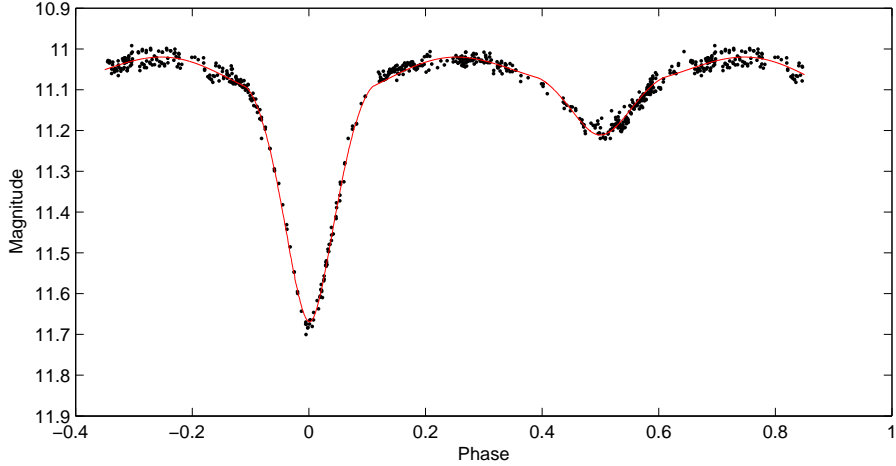}
 \caption{Light curve analysis of RZ~Lyn, based on the Super WASP photometry.}
 \label{FigRZLynLC}
\end{figure}

We carried out an analysis of the LC of this system, using the Super WASP photometry. The results
given in Table \ref{TableLC2} and the plot in Fig. \ref{FigRZLynLC} show that the system contains
one larger component and slightly smaller secondary revolving on mildly inclined orbit. The amount
of the third light is rather negligible.

\begin{figure}
 \includegraphics[width=12cm]{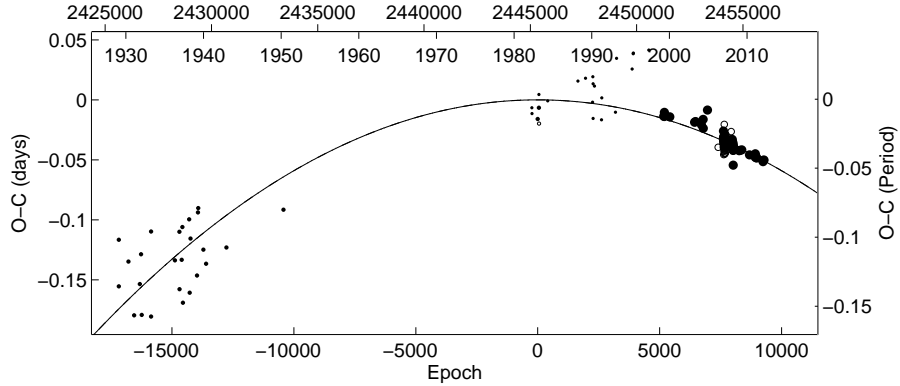}
 \caption{O-C diagram of times of minima for RZ~Lyn.}
 \label{FigRZLynOC}
\end{figure}

For the period analysis we collected the already published data from the $O-C$ gateway and combined
them with our derived minima times from the Super WASP and NSVS photometry. As one can see from our
Fig. \ref{FigRZLynOC}, the system obviously undergoes a mass transfer between the components. Such
an effect is displayed in the $O-C$ diagram as quadratic ephemerides, here in our case the rate is
of about $-5.904 \cdot 10^{-10}$~days.

\subsection{TY~Tri}

The eclipsing binary TY~Tri (also TYC 2312-190-1) was discovered as a variable by
\cite{1963IBVS...21....1W}. No detailed analysis was carried out for this star, only several
publications with the times of minima were published during the last decades. For the light curve
analysis we fixed the primary temperate at a value of 5720~K, in agreement with
\cite{2010PASP..122.1437P} and \cite{2006ApJ...638.1004A}.

\begin{figure}
 \includegraphics[width=12cm]{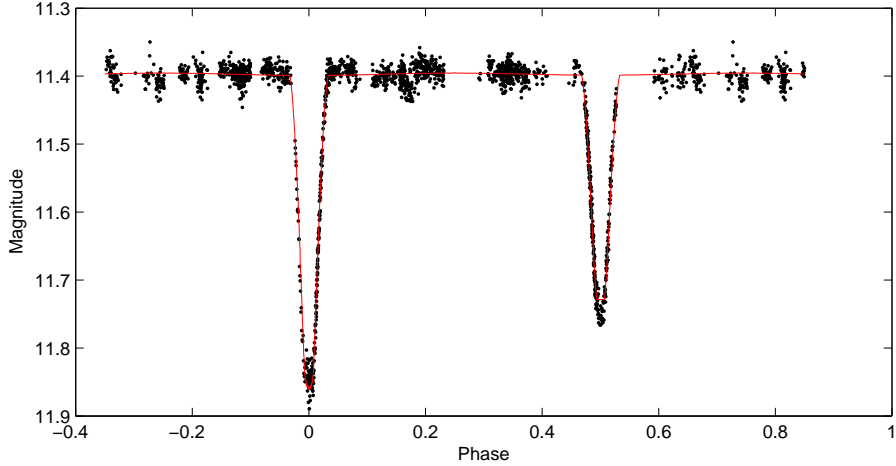}
 \caption{Light curve analysis of TY~Tri, based on the Super WASP photometry.}
 \label{FigTYTriLC}
\end{figure}

\begin{figure}
 \includegraphics[width=12cm]{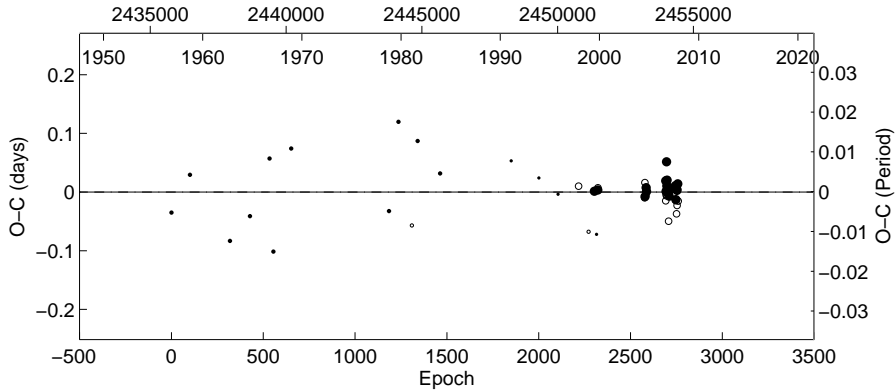}
 \caption{O-C diagram of times of minima for TY~Tri.}
 \label{FigTYTriOC}
\end{figure}

The LC analysis revealed (see Fig. \ref{FigTYTriLC} and Table \ref{TableLC2}) that here we deal
with the most detached system in our sample of stars, but still moving on circular orbit. We see
the system almost exactly edge-on. The third light was not detected. The LC solution was used for
deriving the minima times (both from the Super WASP as well as from the NSVS photometry). The final
plot is given in Fig. \ref{FigTYTriOC}, where we were not able to detect any period variation over
the time span of more than 50~years.

\section{Discussion and conclusions}

The very first LC solution for eight Algol-type eclipsing binaries (based on the Super WASP
photometry) led to several interesting results:

\begin{itemize}
  \item The Super WASP survey served as a unique source of photometric data suitable for the LC
  analysis of many eclipsing binaries never studied before.
  \item The effects of the second order, like the third light, are also detectable in these data.
  \item For two of the systems (EI~Aur, and BP~Dra) the amount of the third light is so large that
  these cannot easily be considered as pure binaries in any future more detailed study.
  \item The method of using the light curve templates for deriving the times of minima provides
  reliable and sufficiently precise minima suitable for a period analysis.
  \item For RZ~Lyn we found a steady period decrease (probably due to mass transfer), while for two
  other systems (XY~Dra and VX~Lac) there were detected the third-body period modulations with
  their respective periods of 18, and 49 years, respectively.
\end{itemize}

All of the presented systems were never been studied before concerning their light curves, hence we
can consider this study as a good starting point for any other future investigators. Especially, a
special focus should be take to these systems, where a larger fraction of the third light was
detected and these systems, where a third body variation in the $O-C$ diagram was detected.

\section{Acknowledgments}
Based on data from the OMC Archive at LAEFF, pre-processed by ISDC. We thank the "ASAS", "NSVS",
"OMC", "CRTS" and "Super WASP" teams for making all of the observations easily public available.
This investigation was supported by the Czech Science Foundation grant no. GA15-02112S. This
research has made use of the SIMBAD database, operated at CDS, Strasbourg, France, and of NASA's
Astrophysics Data System Bibliographic Services.

\newpage

\begin{table}
 \caption{New heliocentric minima times for the studied systems.}
 \label{TableMin1} \centering \scalebox{0.68}{
 \tiny
\begin{tabular}{ c c c c c c | c c c c c c }
\hline \hline
     Star  &  HJD        & Error   & Type & Filter & Source    &       Star  &  HJD        & Error   & Type & Filter & Source   \\
           & 2400000+    & [days]  &      &        &           &             & 2400000+    & [days]  &      &        &          \\ \hline
   EI Aur  & 53239.79843 & 0.00329 & Prim &  W  &  Super WASP  &     XY Dra  & 54648.48213 & 0.00346 & Prim &  W  &  Super WASP \\
   EI Aur  & 53241.63954 & 0.00484 & Sec  &  W  &  Super WASP  &     XY Dra  & 54655.42973 & 0.00442 & Prim &  W  &  Super WASP \\
   EI Aur  & 53249.60626 & 0.00239 & Prim &  W  &  Super WASP  &     XY Dra  & 54662.36851 & 0.00348 & Prim &  W  &  Super WASP \\
   EI Aur  & 53258.81413 & 0.00597 & Sec  &  W  &  Super WASP  &     XY Dra  & 54664.68717 & 0.00709 & Prim &  W  &  Super WASP \\
   EI Aur  & 53260.65985 & 0.00493 & Prim &  W  &  Super WASP  &     XY Dra  & 54669.31447 & 0.00865 & Prim &  W  &  Super WASP \\
   EI Aur  & 53263.72284 & 0.00822 & Sec  &  W  &  Super WASP  &     XY Dra  & 54671.63742 & 0.00484 & Prim &  W  &  Super WASP \\
   EI Aur  & 53271.69447 & 0.00324 & Prim &  W  &  Super WASP  &     XY Dra  & 54676.25627 & 0.00619 & Prim &  W  &  Super WASP \\
   EI Aur  & 53274.75490 & 0.00549 & Sec  &  W  &  Super WASP  &     XY Dra  & 54678.58136 & 0.00540 & Prim &  W  &  Super WASP \\
   EI Aur  & 53276.60882 & 0.00425 & Prim &  W  &  Super WASP  &     XY Dra  & 54685.52612 & 0.00294 & Prim &  W  &  Super WASP \\
   EI Aur  & 54021.81939 & 0.00646 & Sec  &  W  &  Super WASP  &     BP Dra  & 54605.75377 & 0.00331 & Sec  &  W  &  Super WASP \\
   EI Aur  & 54023.65741 & 0.00135 & Prim &  W  &  Super WASP  &     BP Dra  & 54606.74790 & 0.00364 & Sec  &  W  &  Super WASP \\
   EI Aur  & 54031.63289 & 0.00416 & Sec  &  W  &  Super WASP  &     BP Dra  & 54607.74552 & 0.00252 & Sec  &  W  &  Super WASP \\
   EI Aur  & 54050.64116 & 0.00567 & Prim &  W  &  Super WASP  &     BP Dra  & 54608.72035 & 0.00416 & Sec  &  W  &  Super WASP \\
   EI Aur  & 54056.78341 & 0.00574 & Prim &  W  &  Super WASP  &     BP Dra  & 54609.70607 & 0.00376 & Sec  &  W  &  Super WASP \\
   EI Aur  & 54066.59413 & 0.00507 & Prim &  W  &  Super WASP  &     BP Dra  & 54613.65078 & 0.00044 & Sec  &  W  &  Super WASP \\
   EI Aur  & 54068.44404 & 0.00341 & Sec  &  W  &  Super WASP  &     BP Dra  & 54614.64111 & 0.00399 & Sec  &  W  &  Super WASP \\
   EI Aur  & 54069.65880 & 0.00694 & Sec  &  W  &  Super WASP  &     BP Dra  & 54616.61016 & 0.00336 & Sec  &  W  &  Super WASP \\
   EI Aur  & 54071.50796 & 0.00618 & Prim &  W  &  Super WASP  &     BP Dra  & 54618.58910 & 0.00426 & Sec  &  W  &  Super WASP \\
   EI Aur  & 54077.63090 & 0.00380 & Prim &  W  &  Super WASP  &     BP Dra  & 54619.57775 & 0.00327 & Sec  &  W  &  Super WASP \\
   EI Aur  & 54083.77107 & 0.01188 & Prim &  W  &  Super WASP  &     BP Dra  & 54620.56434 & 0.00286 & Sec  &  W  &  Super WASP \\
   EI Aur  & 54085.60965 & 0.00159 & Sec  &  W  &  Super WASP  &     BP Dra  & 54621.54850 & 0.00565 & Sec  &  W  &  Super WASP \\
   EI Aur  & 54087.44571 & 0.00367 & Prim &  W  &  Super WASP  &     BP Dra  & 54622.52987 & 0.00732 & Sec  &  W  &  Super WASP \\
   EI Aur  & 54091.73708 & 0.05763 & Sec  &  W  &  Super WASP  &     BP Dra  & 54623.53236 & 0.00492 & Sec  &  W  &  Super WASP \\
   EI Aur  & 54092.35800 & 0.00456 & Prim &  W  &  Super WASP  &     BP Dra  & 54624.52560 & 0.00456 & Sec  &  W  &  Super WASP \\
   EI Aur  & 54093.57255 & 0.00591 & Prim &  W  &  Super WASP  &     BP Dra  & 54625.49420 & 0.00339 & Sec  &  W  &  Super WASP \\
   EI Aur  & 54109.53910 & 0.00271 & Prim &  W  &  Super WASP  &     BP Dra  & 54626.47436 & 0.00432 & Sec  &  W  &  Super WASP \\
   EI Aur  & 54111.37387 & 0.00574 & Sec  &  W  &  Super WASP  &     BP Dra  & 54639.80828 & 0.00666 & Prim &  W  &  Super WASP \\
   EI Aur  & 54114.43525 & 0.00562 & Prim &  W  &  Super WASP  &     BP Dra  & 54640.77902 & 0.00321 & Prim &  W  &  Super WASP \\
   EI Aur  & 54115.65588 & 0.00512 & Prim &  W  &  Super WASP  &     BP Dra  & 54641.78471 & 0.00518 & Prim &  W  &  Super WASP \\
   EI Aur  & 54116.27000 & 0.00535 & Sec  &  W  &  Super WASP  &     BP Dra  & 54642.76968 & 0.00240 & Prim &  W  &  Super WASP \\
   EI Aur  & 54120.55881 & 0.00533 & Prim &  W  &  Super WASP  &     BP Dra  & 54643.74851 & 0.00445 & Prim &  W  &  Super WASP \\
   EI Aur  & 54122.41118 & 0.00531 & Sec  &  W  &  Super WASP  &     BP Dra  & 54644.74041 & 0.00426 & Prim &  W  &  Super WASP \\
   EI Aur  & 54123.63145 & 0.00675 & Sec  &  W  &  Super WASP  &     BP Dra  & 54645.73031 & 0.00209 & Prim &  W  &  Super WASP \\
   EI Aur  & 54124.24099 & 0.01334 & Prim &  W  &  Super WASP  &     BP Dra  & 54646.72234 & 0.00163 & Prim &  W  &  Super WASP \\
   EI Aur  & 54125.47803 & 0.02013 & Prim &  W  &  Super WASP  &     BP Dra  & 54647.70233 & 0.00203 & Prim &  W  &  Super WASP \\
   EI Aur  & 54135.28435 & 0.00480 & Prim &  W  &  Super WASP  &     BP Dra  & 54649.67851 & 0.00331 & Prim &  W  &  Super WASP \\
   EI Aur  & 54136.51641 & 0.00560 & Prim &  W  &  Super WASP  &     BP Dra  & 54650.66298 & 0.00324 & Prim &  W  &  Super WASP \\
   EI Aur  & 54139.57493 & 0.00679 & Sec  &  W  &  Super WASP  &     BP Dra  & 54651.64956 & 0.00372 & Prim &  W  &  Super WASP \\
   EI Aur  & 54141.42195 & 0.00118 & Prim &  W  &  Super WASP  &     BP Dra  & 54652.63660 & 0.00316 & Prim &  W  &  Super WASP \\
   EI Aur  & 54143.26039 & 0.00200 & Sec  &  W  &  Super WASP  &     BP Dra  & 54655.59807 & 0.00346 & Prim &  W  &  Super WASP \\
   EI Aur  & 54372.65157 & 0.00222 & Sec  &  W  &  Super WASP  &     BP Dra  & 54656.58499 & 0.00065 & Prim &  W  &  Super WASP \\
   EI Aur  & 54381.84920 & 0.00103 & Prim &  W  &  Super WASP  &     BP Dra  & 54657.57141 & 0.00412 & Prim &  W  &  Super WASP \\
   EI Aur  & 54383.69290 & 0.00270 & Sec  &  W  &  Super WASP  &     BP Dra  & 54659.54592 & 0.00355 & Prim &  W  &  Super WASP \\
   EI Aur  & 54388.60045 & 0.00236 & Sec  &  W  &  Super WASP  &     BP Dra  & 54660.53074 & 0.00128 & Prim &  W  &  Super WASP \\
   EI Aur  & 54389.83119 & 0.00707 & Sec  &  W  &  Super WASP  &     BP Dra  & 54661.51919 & 0.00093 & Prim &  W  &  Super WASP \\
   EI Aur  & 54393.50781 & 0.00638 & Sec  &  W  &  Super WASP  &     BP Dra  & 54662.50413 & 0.00138 & Prim &  W  &  Super WASP \\
   EI Aur  & 54394.73397 & 0.00491 & Sec  &  W  &  Super WASP  &     BP Dra  & 54663.48917 & 0.00323 & Prim &  W  &  Super WASP \\
   EI Aur  & 54396.57223 & 0.00222 & Prim &  W  &  Super WASP  &     BP Dra  & 54664.47471 & 0.00485 & Prim &  W  &  Super WASP \\
   EI Aur  & 54399.64381 & 0.00966 & Sec  &  W  &  Super WASP  &     BP Dra  & 54665.46945 & 0.00265 & Prim &  W  &  Super WASP \\
   EI Aur  & 54405.77427 & 0.00712 & Sec  &  W  &  Super WASP  &     BP Dra  & 54666.44639 & 0.00388 & Prim &  W  &  Super WASP \\
   EI Aur  & 54407.61117 & 0.00112 & Prim &  W  &  Super WASP  &     BP Dra  & 54669.41122 & 0.00391 & Prim &  W  &  Super WASP \\
   EI Aur  & 54409.45212 & 0.00863 & Sec  &  W  &  Super WASP  &     BP Dra  & 54670.38242 & 0.00520 & Prim &  W  &  Super WASP \\
   EI Aur  & 54410.68184 & 0.00330 & Sec  &  W  &  Super WASP  &     BP Dra  & 54671.38573 & 0.00531 & Prim &  W  &  Super WASP \\
   EI Aur  & 54418.65078 & 0.00501 & Prim &  W  &  Super WASP  &     BP Dra  & 54673.36105 & 0.00745 & Prim &  W  &  Super WASP \\
   EI Aur  & 54420.49582 & 0.00446 & Sec  &  W  &  Super WASP  &     BP Dra  & 54684.71350 & 0.00374 & Sec  &  W  &  Super WASP \\
   EI Aur  & 54436.44208 & 0.00460 & Sec  &  W  &  Super WASP  &     BP Dra  & 54685.70796 & 0.00596 & Sec  &  W  &  Super WASP \\
   EI Aur  & 54439.50820 & 0.00335 & Prim &  W  &  Super WASP  &     BP Dra  & 54686.67285 & 0.00552 & Sec  &  W  &  Super WASP \\
   EI Aur  & 54447.48593 & 0.03211 & Sec  &  W  &  Super WASP  &     BP Dra  & 54688.65672 & 0.00255 & Sec  &  W  &  Super WASP \\
   EI Aur  & 54452.38034 & 0.00175 & Sec  &  W  &  Super WASP  &     DD Her  & 53159.72406 & 0.00410 & Prim &  W  &  Super WASP \\
   XY Dra  & 54231.74155 & 0.00800 & Prim &  W  &  Super WASP  &     DD Her  & 53165.35782 & 0.00489 & Prim &  W  &  Super WASP \\
   XY Dra  & 54252.57564 & 0.00106 & Prim &  W  &  Super WASP  &     DD Her  & 53171.00389 & 0.02646 & Prim &  W  &  Super WASP \\
   XY Dra  & 54266.47097 & 0.00329 & Prim &  W  &  Super WASP  &     DD Her  & 53176.64755 & 0.00272 & Prim &  W  &  Super WASP \\
   XY Dra  & 54268.78705 & 0.00408 & Prim &  W  &  Super WASP  &     DD Her  & 53182.28829 & 0.00423 & Prim &  W  &  Super WASP \\
   XY Dra  & 54273.41399 & 0.01264 & Prim &  W  &  Super WASP  &     DD Her  & 53193.57817 & 0.00862 & Prim &  W  &  Super WASP \\
   XY Dra  & 54275.72631 & 0.00364 & Prim &  W  &  Super WASP  &     DD Her  & 53199.21100 & 0.01450 & Prim &  W  &  Super WASP \\
   XY Dra  & 54280.36463 & 0.00492 & Prim &  W  &  Super WASP  &     DD Her  & 53204.86915 & 0.00085 & Prim &  W  &  Super WASP \\
   XY Dra  & 54282.67744 & 0.00736 & Prim &  W  &  Super WASP  &     DD Her  & 53221.80448 & 0.00422 & Prim &  W  &  Super WASP \\
   XY Dra  & 54287.30232 & 0.00328 & Prim &  W  &  Super WASP  &     DD Her  & 53227.44355 & 0.00011 & Prim &  W  &  Super WASP \\
   XY Dra  & 54289.61817 & 0.00214 & Prim &  W  &  Super WASP  &     DD Her  & 53238.72048 & 0.00310 & Prim &  W  &  Super WASP \\
   XY Dra  & 54296.56650 & 0.00445 & Prim &  W  &  Super WASP  &     DD Her  & 53261.29036 & 0.00998 & Prim &  W  &  Super WASP \\
   XY Dra  & 54303.51362 & 0.00506 & Prim &  W  &  Super WASP  &     DD Her  & 53278.23823 & 0.06578 & Prim &  W  &  Super WASP \\
   XY Dra  & 54317.41155 & 0.00708 & Prim &  W  &  Super WASP  &     DD Her  & 52956.54640 & 0.01290 & Prim &  V  &  ASAS       \\
   XY Dra  & 54324.33750 & 0.00906 & Prim &  W  &  Super WASP  &     DD Her  & 51449.76387 & 0.00529 & Prim &  C  &  NSVS       \\
   XY Dra  & 54333.60940 & 0.00314 & Prim &  W  &  Super WASP  &     VX Lac  & 53164.64384 & 0.00064 & Prim &  W  &  Super WASP \\
   XY Dra  & 54576.70876 & 0.00771 & Prim &  W  &  Super WASP  &     VX Lac  & 53165.71872 & 0.00093 & Prim &  W  &  Super WASP \\
   XY Dra  & 54597.54636 & 0.00189 & Prim &  W  &  Super WASP  &     VX Lac  & 53177.54044 & 0.00093 & Prim &  W  &  Super WASP \\
   XY Dra  & 54606.81787 & 0.00920 & Prim &  W  &  Super WASP  &     VX Lac  & 53178.61181 & 0.00502 & Prim &  W  &  Super WASP \\
   XY Dra  & 54613.75888 & 0.00567 & Prim &  W  &  Super WASP  &     VX Lac  & 53179.68697 & 0.00016 & Prim &  W  &  Super WASP \\
   XY Dra  & 54618.38035 & 0.00658 & Prim &  W  &  Super WASP  &     VX Lac  & 53180.76028 & 0.00367 & Prim &  W  &  Super WASP \\
   XY Dra  & 54620.69641 & 0.00542 & Prim &  W  &  Super WASP  &     VX Lac  & 53191.50581 & 0.00255 & Prim &  W  &  Super WASP \\
   XY Dra  & 54627.64153 & 0.00286 & Prim &  W  &  Super WASP  &     VX Lac  & 53192.58112 & 0.00257 & Prim &  W  &  Super WASP \\
   XY Dra  & 54636.89985 & 0.00560 & Prim &  W  &  Super WASP  &     VX Lac  & 53193.65503 & 0.00444 & Prim &  W  &  Super WASP \\
   XY Dra  & 54641.53766 & 0.00398 & Prim &  W  &  Super WASP  &     VX Lac  & 53194.72972 & 0.00727 & Prim &  W  &  Super WASP \\
 \hline
\end{tabular}}
\end{table}

\newpage

\begin{table}
 \caption{New heliocentric minima times for the studied systems - cont.1}
 \label{TableMin2} \centering \scalebox{0.68}{
 \tiny
\begin{tabular}{ c c c c c c | c c c c c c }
\hline \hline
     Star  &  HJD        & Error   & Type & Filter & Source    &      Star  &  HJD        & Error   & Type & Filter & Source     \\
           & 2400000+    & [days]  &      &        &           &            & 2400000+    & [days]  &      &        &            \\ \hline
   VX Lac  & 53205.47578 & 0.00122 & Prim &  W  &  Super WASP  &    WX Lib  & 54203.45147 & 0.00373 & Sec  &  W  &  Super WASP   \\
   VX Lac  & 53206.54973 & 0.00028 & Prim &  W  &  Super WASP  &    WX Lib  & 54207.59244 & 0.00055 & Prim &  W  &  Super WASP   \\
   VX Lac  & 53207.62366 & 0.00000 & Prim &  W  &  Super WASP  &    WX Lib  & 54211.27564 & 0.00301 & Prim &  W  &  Super WASP   \\
   VX Lac  & 53208.69909 & 0.00187 & Prim &  W  &  Super WASP  &    WX Lib  & 54212.65320 & 0.00299 & Sec  &  W  &  Super WASP   \\
   VX Lac  & 53209.77090 & 0.00566 & Prim &  W  &  Super WASP  &    WX Lib  & 54213.57224 & 0.00253 & Sec  &  W  &  Super WASP   \\
   VX Lac  & 53219.44361 & 0.01045 & Prim &  W  &  Super WASP  &    WX Lib  & 54214.49680 & 0.00271 & Sec  &  W  &  Super WASP   \\
   VX Lac  & 53220.51806 & 0.00044 & Prim &  W  &  Super WASP  &    WX Lib  & 54215.41445 & 0.00190 & Sec  &  W  &  Super WASP   \\
   VX Lac  & 53222.66684 & 0.00256 & Prim &  W  &  Super WASP  &    WX Lib  & 54216.33211 & 0.00342 & Sec  &  W  &  Super WASP   \\
   VX Lac  & 53223.73986 & 0.00171 & Prim &  W  &  Super WASP  &    WX Lib  & 54230.59395 & 0.00334 & Prim &  W  &  Super WASP   \\
   VX Lac  & 53235.56067 & 0.00081 & Prim &  W  &  Super WASP  &    WX Lib  & 54231.51311 & 0.00368 & Prim &  W  &  Super WASP   \\
   VX Lac  & 53236.63663 & 0.00340 & Prim &  W  &  Super WASP  &    WX Lib  & 54232.43401 & 0.00175 & Prim &  W  &  Super WASP   \\
   VX Lac  & 53247.38067 & 0.00225 & Prim &  W  &  Super WASP  &    WX Lib  & 54233.35299 & 0.00100 & Prim &  W  &  Super WASP   \\
   VX Lac  & 53248.45479 & 0.00574 & Prim &  W  &  Super WASP  &    WX Lib  & 54236.57262 & 0.00463 & Sec  &  W  &  Super WASP   \\
   VX Lac  & 53249.52955 & 0.00061 & Prim &  W  &  Super WASP  &    WX Lib  & 54237.49040 & 0.00451 & Sec  &  W  &  Super WASP   \\
   VX Lac  & 53250.60386 & 0.00269 & Prim &  W  &  Super WASP  &    WX Lib  & 54238.41622 & 0.00475 & Sec  &  W  &  Super WASP   \\
   VX Lac  & 53252.75332 & 0.00219 & Prim &  W  &  Super WASP  &    WX Lib  & 54239.33132 & 0.00240 & Sec  &  W  &  Super WASP   \\
   VX Lac  & 53261.35027 & 0.00690 & Prim &  W  &  Super WASP  &    WX Lib  & 54244.39194 & 0.00406 & Prim &  W  &  Super WASP   \\
   VX Lac  & 53262.42319 & 0.00391 & Prim &  W  &  Super WASP  &    WX Lib  & 54246.23718 & 0.00341 & Prim &  W  &  Super WASP   \\
   VX Lac  & 53263.49786 & 0.00306 & Prim &  W  &  Super WASP  &    WX Lib  & 54254.51637 & 0.00256 & Prim &  W  &  Super WASP   \\
   VX Lac  & 53264.57568 & 0.00042 & Prim &  W  &  Super WASP  &    WX Lib  & 54257.27174 & 0.00161 & Prim &  W  &  Super WASP   \\
   VX Lac  & 53265.64647 & 0.00178 & Prim &  W  &  Super WASP  &    WX Lib  & 54268.31725 & 0.00253 & Prim &  W  &  Super WASP   \\
   VX Lac  & 53266.72288 & 0.00514 & Prim &  W  &  Super WASP  &    WX Lib  & 54269.23159 & 0.00974 & Prim &  W  &  Super WASP   \\
   VX Lac  & 53275.31899 & 0.00179 & Prim &  W  &  Super WASP  &    WX Lib  & 54271.53030 & 0.00342 & Sec  &  W  &  Super WASP   \\
   VX Lac  & 53276.39168 & 0.00146 & Prim &  W  &  Super WASP  &    WX Lib  & 54272.45350 & 0.00026 & Sec  &  W  &  Super WASP   \\
   VX Lac  & 53277.46644 & 0.00017 & Prim &  W  &  Super WASP  &    WX Lib  & 54273.37246 & 0.00230 & Sec  &  W  &  Super WASP   \\
   VX Lac  & 53278.54135 & 0.00128 & Prim &  W  &  Super WASP  &    WX Lib  & 54274.29395 & 0.00122 & Sec  &  W  &  Super WASP   \\
   VX Lac  & 53942.58409 & 0.00119 & Prim &  W  &  Super WASP  &    WX Lib  & 54287.17233 & 0.00722 & Sec  &  W  &  Super WASP   \\
   VX Lac  & 53943.65925 & 0.00521 & Prim &  W  &  Super WASP  &    WX Lib  & 54291.31331 & 0.00416 & Prim &  W  &  Super WASP   \\
   VX Lac  & 53944.73254 & 0.00074 & Prim &  W  &  Super WASP  &    WX Lib  & 54292.23335 & 0.00323 & Prim &  W  &  Super WASP   \\
   VX Lac  & 53955.47843 & 0.00100 & Prim &  W  &  Super WASP  &    WX Lib  & 54295.45465 & 0.00216 & Sec  &  W  &  Super WASP   \\
   VX Lac  & 53969.44742 & 0.00011 & Prim &  W  &  Super WASP  &    WX Lib  & 54296.37446 & 0.00315 & Sec  &  W  &  Super WASP   \\
   VX Lac  & 53970.52161 & 0.00018 & Prim &  W  &  Super WASP  &    WX Lib  & 54297.29830 & 0.00350 & Sec  &  W  &  Super WASP   \\
   VX Lac  & 53971.59617 & 0.00143 & Prim &  W  &  Super WASP  &    WX Lib  & 54298.21067 & 0.00324 & Sec  &  W  &  Super WASP   \\
   VX Lac  & 53972.67090 & 0.00867 & Prim &  W  &  Super WASP  &    WX Lib  & 54301.43240 & 0.00495 & Prim &  W  &  Super WASP   \\
   VX Lac  & 53973.74376 & 0.01138 & Prim &  W  &  Super WASP  &    WX Lib  & 54337.31192 & 0.00837 & Prim &  W  &  Super WASP   \\
   VX Lac  & 53997.38411 & 0.00155 & Prim &  W  &  Super WASP  &    WX Lib  & 54338.23273 & 0.00497 & Prim &  W  &  Super WASP   \\
   VX Lac  & 53998.45861 & 0.00176 & Prim &  W  &  Super WASP  &    WX Lib  & 54518.55510 & 0.00061 & Prim &  W  &  Super WASP   \\
   VX Lac  & 54056.48221 & 0.00697 & Prim &  W  &  Super WASP  &    WX Lib  & 54525.45469 & 0.00001 & Sec  &  W  &  Super WASP   \\
   VX Lac  & 54057.55518 & 0.01641 & Prim &  W  &  Super WASP  &    WX Lib  & 54528.67506 & 0.00163 & Prim &  W  &  Super WASP   \\
   VX Lac  & 54303.61812 & 0.00137 & Prim &  W  &  Super WASP  &    WX Lib  & 54536.49703 & 0.00071 & Sec  &  W  &  Super WASP   \\
   VX Lac  & 54304.69283 & 0.00164 & Prim &  W  &  Super WASP  &    WX Lib  & 54547.53886 & 0.00359 & Sec  &  W  &  Super WASP   \\
   VX Lac  & 54315.44317 & 0.00666 & Prim &  W  &  Super WASP  &    WX Lib  & 54553.51620 & 0.00018 & Prim &  W  &  Super WASP   \\
   VX Lac  & 54316.51092 & 0.00706 & Prim &  W  &  Super WASP  &    WX Lib  & 54554.43505 & 0.00244 & Prim &  W  &  Super WASP   \\
   VX Lac  & 54318.66166 & 0.00080 & Prim &  W  &  Super WASP  &    WX Lib  & 54555.35670 & 0.00010 & Prim &  W  &  Super WASP   \\
   VX Lac  & 54319.73556 & 0.00144 & Prim &  W  &  Super WASP  &    WX Lib  & 54557.65679 & 0.00086 & Sec  &  W  &  Super WASP   \\
   VX Lac  & 54329.41159 & 0.00045 & Prim &  W  &  Super WASP  &    WX Lib  & 54558.57649 & 0.00009 & Sec  &  W  &  Super WASP   \\
   VX Lac  & 54330.48057 & 0.00073 & Prim &  W  &  Super WASP  &    WX Lib  & 54560.41698 & 0.00049 & Sec  &  W  &  Super WASP   \\
   VX Lac  & 54331.55560 & 0.00012 & Prim &  W  &  Super WASP  &    WX Lib  & 54561.33652 & 0.00325 & Sec  &  W  &  Super WASP   \\
   VX Lac  & 54332.62991 & 0.00030 & Prim &  W  &  Super WASP  &    WX Lib  & 54564.55555 & 0.00134 & Prim &  W  &  Super WASP   \\
   VX Lac  & 54333.70456 & 0.00274 & Prim &  W  &  Super WASP  &    WX Lib  & 54565.47622 & 0.00016 & Prim &  W  &  Super WASP   \\
   VX Lac  & 54334.77826 & 0.00490 & Prim &  W  &  Super WASP  &    WX Lib  & 54566.39604 & 0.00054 & Prim &  W  &  Super WASP   \\
   VX Lac  & 54344.44749 & 0.00341 & Prim &  W  &  Super WASP  &    WX Lib  & 54571.45482 & 0.00103 & Sec  &  W  &  Super WASP   \\
   VX Lac  & 54345.52456 & 0.00060 & Prim &  W  &  Super WASP  &    WX Lib  & 54572.37522 & 0.00028 & Sec  &  W  &  Super WASP   \\
   VX Lac  & 54346.59859 & 0.00158 & Prim &  W  &  Super WASP  &    WX Lib  & 54573.29495 & 0.00006 & Sec  &  W  &  Super WASP   \\
   VX Lac  & 54347.67305 & 0.00069 & Prim &  W  &  Super WASP  &    WX Lib  & 54574.67423 & 0.00154 & Prim &  W  &  Super WASP   \\
   VX Lac  & 54348.74674 & 0.00272 & Prim &  W  &  Super WASP  &    WX Lib  & 54581.57478 & 0.00042 & Sec  &  W  &  Super WASP   \\
   VX Lac  & 54357.34618 & 0.00352 & Prim &  W  &  Super WASP  &    WX Lib  & 54582.49226 & 0.00125 & Sec  &  W  &  Super WASP   \\
   VX Lac  & 54358.41811 & 0.00063 & Prim &  W  &  Super WASP  &    WX Lib  & 54583.41896 & 0.00113 & Sec  &  W  &  Super WASP   \\
   VX Lac  & 54359.49267 & 0.00038 & Prim &  W  &  Super WASP  &    WX Lib  & 54584.33531 & 0.00174 & Sec  &  W  &  Super WASP   \\
   VX Lac  & 54360.56791 & 0.00177 & Prim &  W  &  Super WASP  &    WX Lib  & 54586.63520 & 0.00011 & Prim &  W  &  Super WASP   \\
   VX Lac  & 54361.64171 & 0.00008 & Prim &  W  &  Super WASP  &    WX Lib  & 54588.47344 & 0.00010 & Prim &  W  &  Super WASP   \\
   VX Lac  & 54362.71659 & 0.00182 & Prim &  W  &  Super WASP  &    WX Lib  & 54592.61361 & 0.00380 & Sec  &  W  &  Super WASP   \\
   VX Lac  & 54372.38510 & 0.00065 & Prim &  W  &  Super WASP  &    WX Lib  & 54594.45681 & 0.00167 & Sec  &  W  &  Super WASP   \\
   VX Lac  & 54373.46159 & 0.00071 & Prim &  W  &  Super WASP  &    WX Lib  & 54600.43937 & 0.00109 & Prim &  W  &  Super WASP   \\
   VX Lac  & 54374.53624 & 0.00228 & Prim &  W  &  Super WASP  &    WX Lib  & 54601.35945 & 0.00031 & Prim &  W  &  Super WASP   \\
   VX Lac  & 54387.43025 & 0.00007 & Prim &  W  &  Super WASP  &    WX Lib  & 54609.17518 & 0.00461 & Sec  &  W  &  Super WASP   \\
   VX Lac  & 54388.50380 & 0.00106 & Prim &  W  &  Super WASP  &    WX Lib  & 54612.39907 & 0.00464 & Prim &  W  &  Super WASP   \\
   VX Lac  & 54389.57750 & 0.00683 & Prim &  W  &  Super WASP  &    WX Lib  & 54614.23742 & 0.00184 & Prim &  W  &  Super WASP   \\
   VX Lac  & 54399.25015 & 0.00494 & Prim &  W  &  Super WASP  &    WX Lib  & 52214.85796 & 0.00135 & Prim &  V  &  ASAS         \\
   VX Lac  & 54402.47315 & 0.00069 & Prim &  W  &  Super WASP  &    WX Lib  & 52215.31522 & 0.00056 & Sec  &  V  &  ASAS         \\
   VX Lac  & 54416.43920 & 0.00156 & Prim &  W  &  Super WASP  &    WX Lib  & 52770.54040 & 0.00101 & Prim &  V  &  ASAS         \\
   VX Lac  & 51493.78293 & 0.00245 & Prim &  C  &  NSVS        &    WX Lib  & 52771.00118 & 0.00140 & Sec  &  V  &  ASAS         \\
   VX Lac  & 51494.32115 & 0.00452 & Prim &  C  &  NSVS        &    WX Lib  & 53143.14430 & 0.00133 & Prim &  V  &  ASAS         \\
   WX Lib  & 54151.47369 & 0.00134 & Prim &  W  &  Super WASP  &    WX Lib  & 53143.60384 & 0.00246 & Sec  &  V  &  ASAS         \\
   WX Lib  & 54155.61096 & 0.00022 & Sec  &  W  &  Super WASP  &    WX Lib  & 53510.22928 & 0.00343 & Prim &  V  &  ASAS         \\
   WX Lib  & 54156.53318 & 0.00088 & Sec  &  W  &  Super WASP  &    WX Lib  & 53510.68764 & 0.00091 & Sec  &  V  &  ASAS         \\
   WX Lib  & 54175.39377 & 0.00056 & Prim &  W  &  Super WASP  &    WX Lib  & 53824.86673 & 0.00125 & Prim &  V  &  ASAS         \\
   WX Lib  & 54179.53406 & 0.00129 & Sec  &  W  &  Super WASP  &    WX Lib  & 53825.32650 & 0.00139 & Sec  &  V  &  ASAS         \\
   WX Lib  & 54180.45331 & 0.00313 & Sec  &  W  &  Super WASP  &    WX Lib  & 54239.79485 & 0.00092 & Prim &  V  &  ASAS         \\
   WX Lib  & 54181.37149 & 0.00156 & Sec  &  W  &  Super WASP  &    WX Lib  & 54240.24947 & 0.00289 & Sec  &  V  &  ASAS         \\
   WX Lib  & 54186.43204 & 0.00817 & Prim &  W  &  Super WASP  &    WX Lib  & 54607.79578 & 0.00083 & Prim &  V  &  ASAS         \\
   WX Lib  & 54199.31531 & 0.00019 & Prim &  W  &  Super WASP  &    WX Lib  & 54608.25407 & 0.00079 & Sec  &  V  &  ASAS         \\
   WX Lib  & 54200.69220 & 0.00026 & Sec  &  W  &  Super WASP  &    WX Lib  & 54973.03732 & 0.00069 & Prim &  V  &  ASAS         \\
 \hline
\end{tabular}}
\end{table}

\newpage

\begin{table}
 \caption{New heliocentric minima times for the studied systems - cont.2}
 \label{TableMin3} \centering \scalebox{0.68}{
 \tiny
\begin{tabular}{ c c c c c c | c c c c c c }
\hline \hline
     Star  &  HJD        & Error   & Type & Filter & Source   &     Star  &  HJD        & Error   & Type & Filter & Source    \\
           & 2400000+    & [days]  &      &        &          &           & 2400000+    & [days]  &      &        &           \\ \hline
   WX Lib  & 54973.49691 & 0.00168 & Sec  &  V  &  ASAS       &   RZ Lyn  & 54530.62985 & 0.00208 & Prim &  W  &  Super WASP  \\
   WX Lib  & 53605.90992 & 0.00158 & Prim &  V  &  CRTS       &   RZ Lyn  & 54532.34972 & 0.00083 & Sec  &  W  &  Super WASP  \\
   WX Lib  & 53821.18676 & 0.00528 & Prim &  V  &  CRTS       &   RZ Lyn  & 54533.49086 & 0.00192 & Sec  &  W  &  Super WASP  \\
   WX Lib  & 53832.22397 & 0.00454 & Prim &  V  &  CRTS       &   RZ Lyn  & 54535.21150 & 0.00566 & Prim &  W  &  Super WASP  \\
   WX Lib  & 53855.22189 & 0.00180 & Prim &  V  &  CRTS       &   RZ Lyn  & 54536.36128 & 0.00184 & Prim &  W  &  Super WASP  \\
   WX Lib  & 53873.16292 & 0.01015 & Sec  &  V  &  CRTS       &   RZ Lyn  & 54539.22670 & 0.00081 & Sec  &  W  &  Super WASP  \\
   WX Lib  & 53878.22712 & 0.07436 & Prim &  V  &  CRTS       &   RZ Lyn  & 54544.38890 & 0.00211 & Prim &  W  &  Super WASP  \\
   WX Lib  & 53886.05374 & 0.00578 & Sec  &  V  &  CRTS       &   RZ Lyn  & 54545.52136 & 0.00678 & Prim &  W  &  Super WASP  \\
   WX Lib  & 53914.10772 & 0.00011 & Prim &  V  &  CRTS       &   RZ Lyn  & 54553.56534 & 0.00292 & Prim &  W  &  Super WASP  \\
   WX Lib  & 53920.08515 & 0.00390 & Sec  &  V  &  CRTS       &   RZ Lyn  & 54555.28354 & 0.00339 & Sec  &  W  &  Super WASP  \\
   WX Lib  & 54120.19624 & 0.01175 & Prim &  V  &  CRTS       &   RZ Lyn  & 54556.43655 & 0.00468 & Sec  &  W  &  Super WASP  \\
   WX Lib  & 54188.26766 & 0.00072 & Prim &  V  &  CRTS       &   RZ Lyn  & 54557.57995 & 0.00276 & Sec  &  W  &  Super WASP  \\
   WX Lib  & 54237.02867 & 0.00511 & Prim &  V  &  CRTS       &   RZ Lyn  & 54559.29697 & 0.00325 & Prim &  W  &  Super WASP  \\
   WX Lib  & 54266.01160 & 0.00393 & Sec  &  V  &  CRTS       &   RZ Lyn  & 51305.54688 & 0.00099 & Prim &  C  &  NSVS        \\
   WX Lib  & 54318.91224 & 0.06814 & Prim &  V  &  CRTS       &   RZ Lyn  & 51306.11615 & 0.00892 & Prim &  C  &  NSVS        \\
   WX Lib  & 54651.95046 & 0.02253 & Prim &  V  &  CRTS       &   RZ Lyn  & 51562.45080 & 0.00146 & Sec  &  C  &  NSVS        \\
   WX Lib  & 54861.26160 & 0.00586 & Sec  &  V  &  CRTS       &   RZ Lyn  & 51563.02362 & 0.00657 & Prim &  C  &  NSVS        \\
   WX Lib  & 55743.07995 & 0.00280 & Prim &  V  &  CRTS       &   TY Tri  & 53200.88439 & 0.02149 & Sec  &  W  &  Super WASP  \\
   WX Lib  & 56029.20766 & 0.00551 & Prim &  V  &  CRTS       &   TY Tri  & 53214.43429 & 0.00670 & Sec  &  W  &  Super WASP  \\
   WX Lib  & 53765.07400 & 0.00339 & Prim &  V  &  OMC        &   TY Tri  & 53217.79158 & 0.00986 & Prim &  W  &  Super WASP  \\
   WX Lib  & 53765.53210 & 0.00305 & Sec  &  V  &  OMC        &   TY Tri  & 53241.46097 & 0.00276 & Sec  &  W  &  Super WASP  \\
   WX Lib  & 53765.99313 & 0.00367 & Prim &  V  &  OMC        &   TY Tri  & 53258.37890 & 0.00543 & Prim &  W  &  Super WASP  \\
   WX Lib  & 53768.30057 & 0.00439 & Sec  &  V  &  OMC        &   TY Tri  & 53261.75425 & 0.02789 & Sec  &  W  &  Super WASP  \\
   WX Lib  & 53770.59579 & 0.00515 & Prim &  V  &  OMC        &   TY Tri  & 53271.89645 & 0.00806 & Prim &  W  &  Super WASP  \\
   WX Lib  & 55619.79983 & 0.00123 & Prim &  V  &  OMC        &   TY Tri  & 53971.76602 & 0.00916 & Sec  &  W  &  Super WASP  \\
   WX Lib  & 55624.40569 & 0.00575 & Prim &  V  &  OMC        &   TY Tri  & 53978.51322 & 0.01864 & Sec  &  W  &  Super WASP  \\
   WX Lib  & 55978.61573 & 0.00202 & Prim &  V  &  OMC        &   TY Tri  & 53981.92835 & 0.02444 & Prim &  W  &  Super WASP  \\
   WX Lib  & 55980.91124 & 0.00224 & Sec  &  V  &  OMC        &   TY Tri  & 53995.43489 & 0.01649 & Prim &  W  &  Super WASP  \\
   WX Lib  & 55982.29267 & 0.00385 & Prim &  V  &  OMC        &   TY Tri  & 53998.80999 & 0.00447 & Sec  &  W  &  Super WASP  \\
   WX Lib  & 55983.67560 & 0.00313 & Sec  &  V  &  OMC        &   TY Tri  & 54005.57006 & 0.00086 & Sec  &  W  &  Super WASP  \\
   RZ Lyn  & 53131.42044 & 0.00416 & Prim &  W  &  Super WASP &   TY Tri  & 54009.00902 & 0.00579 & Prim &  W  &  Super WASP  \\
   RZ Lyn  & 53132.55995 & 0.00572 & Prim &  W  &  Super WASP &   TY Tri  & 54022.49202 & 0.00866 & Prim &  W  &  Super WASP  \\
   RZ Lyn  & 53837.31985 & 0.00536 & Sec  &  W  &  Super WASP &   TY Tri  & 54029.26387 & 0.01225 & Prim &  W  &  Super WASP  \\
   RZ Lyn  & 54084.49234 & 0.00285 & Prim &  W  &  Super WASP &   TY Tri  & 54049.53258 & 0.00363 & Prim &  W  &  Super WASP  \\
   RZ Lyn  & 54085.63346 & 0.00151 & Prim &  W  &  Super WASP &   TY Tri  & 54056.29640 & 0.01190 & Prim &  W  &  Super WASP  \\
   RZ Lyn  & 54092.51240 & 0.00421 & Prim &  W  &  Super WASP &   TY Tri  & 54066.43039 & 0.00868 & Sec  &  W  &  Super WASP  \\
   RZ Lyn  & 54094.80424 & 0.00537 & Prim &  W  &  Super WASP &   TY Tri  & 54076.57522 & 0.00655 & Prim &  W  &  Super WASP  \\
   RZ Lyn  & 54098.80906 & 0.00444 & Sec  &  W  &  Super WASP &   TY Tri  & 54083.34766 & 0.00495 & Prim &  W  &  Super WASP  \\
   RZ Lyn  & 54100.54461 & 0.00182 & Prim &  W  &  Super WASP &   TY Tri  & 54086.67109 & 0.00998 & Sec  &  W  &  Super WASP  \\
   RZ Lyn  & 54101.68912 & 0.00028 & Prim &  W  &  Super WASP &   TY Tri  & 54093.48381 & 0.00552 & Sec  &  W  &  Super WASP  \\
   RZ Lyn  & 54109.71604 & 0.00337 & Prim &  W  &  Super WASP &   TY Tri  & 54103.61895 & 0.00645 & Prim &  W  &  Super WASP  \\
   RZ Lyn  & 54114.86844 & 0.00616 & Sec  &  W  &  Super WASP &   TY Tri  & 54333.54585 & 0.00241 & Prim &  W  &  Super WASP  \\
   RZ Lyn  & 54115.45404 & 0.00586 & Prim &  W  &  Super WASP &   TY Tri  & 54350.42559 & 0.00102 & Sec  &  W  &  Super WASP  \\
   RZ Lyn  & 54118.88614 & 0.00397 & Prim &  W  &  Super WASP &   TY Tri  & 54353.80826 & 0.00950 & Prim &  W  &  Super WASP  \\
   RZ Lyn  & 54120.62547 & 0.00184 & Sec  &  W  &  Super WASP &   TY Tri  & 54360.58907 & 0.00525 & Prim &  W  &  Super WASP  \\
   RZ Lyn  & 54121.75079 & 0.00086 & Sec  &  W  &  Super WASP &   TY Tri  & 54377.45125 & 0.05284 & Sec  &  W  &  Super WASP  \\
   RZ Lyn  & 54123.48066 & 0.00462 & Prim &  W  &  Super WASP &   TY Tri  & 54387.63625 & 0.00354 & Prim &  W  &  Super WASP  \\
   RZ Lyn  & 54139.53995 & 0.00173 & Prim &  W  &  Super WASP &   TY Tri  & 54394.39614 & 0.00032 & Prim &  W  &  Super WASP  \\
   RZ Lyn  & 54140.68419 & 0.00173 & Prim &  W  &  Super WASP &   TY Tri  & 54397.75156 & 0.01359 & Sec  &  W  &  Super WASP  \\
   RZ Lyn  & 54141.82840 & 0.00538 & Prim &  W  &  Super WASP &   TY Tri  & 54421.45550 & 0.00405 & Prim &  W  &  Super WASP  \\
   RZ Lyn  & 54142.39263 & 0.00234 & Sec  &  W  &  Super WASP &   TY Tri  & 54438.33122 & 0.00176 & Sec  &  W  &  Super WASP  \\
   RZ Lyn  & 54143.53890 & 0.00265 & Sec  &  W  &  Super WASP &   TY Tri  & 51479.95978 & 0.00255 & Prim &  C  &  NSVS        \\
   RZ Lyn  & 54146.41102 & 0.00220 & Prim &  W  &  Super WASP &   TY Tri  & 51483.34452 & 0.00193 & Sec  &  C  &  NSVS        \\
   RZ Lyn  & 54147.56655 & 0.00283 & Prim &  W  &  Super WASP &           &             &         &      &     &              \\
   RZ Lyn  & 54150.42389 & 0.00427 & Sec  &  W  &  Super WASP &           &             &         &      &     &              \\
   RZ Lyn  & 54152.71500 & 0.00285 & Sec  &  W  &  Super WASP &           &             &         &      &     &              \\
   RZ Lyn  & 54154.44729 & 0.00008 & Prim &  W  &  Super WASP &           &             &         &      &     &              \\
   RZ Lyn  & 54155.59752 & 0.00053 & Prim &  W  &  Super WASP &           &             &         &      &     &              \\
   RZ Lyn  & 54156.73595 & 0.00002 & Prim &  W  &  Super WASP &           &             &         &      &     &              \\
   RZ Lyn  & 54157.31486 & 0.00420 & Sec  &  W  &  Super WASP &           &             &         &      &     &              \\
   RZ Lyn  & 54158.45739 & 0.00507 & Sec  &  W  &  Super WASP &           &             &         &      &     &              \\
   RZ Lyn  & 54159.61127 & 0.00105 & Sec  &  W  &  Super WASP &           &             &         &      &     &              \\
   RZ Lyn  & 54165.33117 & 0.00181 & Sec  &  W  &  Super WASP &           &             &         &      &     &              \\
   RZ Lyn  & 54166.49159 & 0.00318 & Sec  &  W  &  Super WASP &           &             &         &      &     &              \\
   RZ Lyn  & 54167.63178 & 0.00264 & Sec  &  W  &  Super WASP &           &             &         &      &     &              \\
   RZ Lyn  & 54169.35188 & 0.00120 & Prim &  W  &  Super WASP &           &             &         &      &     &              \\
   RZ Lyn  & 54170.50493 & 0.00312 & Prim &  W  &  Super WASP &           &             &         &      &     &              \\
   RZ Lyn  & 54171.64906 & 0.00043 & Prim &  W  &  Super WASP &           &             &         &      &     &              \\
   RZ Lyn  & 54192.29469 & 0.00138 & Prim &  W  &  Super WASP &           &             &         &      &     &              \\
   RZ Lyn  & 54194.58713 & 0.00269 & Prim &  W  &  Super WASP &           &             &         &      &     &              \\
   RZ Lyn  & 54202.61239 & 0.00103 & Prim &  W  &  Super WASP &           &             &         &      &     &              \\
   RZ Lyn  & 54205.47774 & 0.00374 & Sec  &  W  &  Super WASP &           &             &         &      &     &              \\
   RZ Lyn  & 54206.62493 & 0.00497 & Sec  &  W  &  Super WASP &           &             &         &      &     &              \\
   RZ Lyn  & 54208.34604 & 0.00135 & Prim &  W  &  Super WASP &           &             &         &      &     &              \\
   RZ Lyn  & 54210.64324 & 0.00194 & Prim &  W  &  Super WASP &           &             &         &      &     &              \\
   RZ Lyn  & 54213.50360 & 0.00322 & Sec  &  W  &  Super WASP &           &             &         &      &     &              \\
   RZ Lyn  & 54436.58182 & 0.00238 & Prim &  W  &  Super WASP &           &             &         &      &     &              \\
   RZ Lyn  & 54437.73058 & 0.00373 & Prim &  W  &  Super WASP &           &             &         &      &     &              \\
   RZ Lyn  & 54438.88004 & 0.00498 & Prim &  W  &  Super WASP &           &             &         &      &     &              \\
   RZ Lyn  & 54439.45100 & 0.00514 & Sec  &  W  &  Super WASP &           &             &         &      &     &              \\
   RZ Lyn  & 54447.48858 & 0.00467 & Sec  &  W  &  Super WASP &           &             &         &      &     &              \\
   RZ Lyn  & 54497.37261 & 0.00198 & Prim &  W  &  Super WASP &           &             &         &      &     &              \\
   RZ Lyn  & 54501.38003 & 0.00434 & Sec  &  W  &  Super WASP &           &             &         &      &     &              \\
 \hline
\end{tabular}}
\end{table}

\end{document}